\documentclass[aps,pra,preprint,groupedaddress,tightenlines]{revtex4}

\usepackage{amsmath}
\usepackage[english]{babel}

\def\bra#1{{\langle#1|}}
\def\ket#1{{|#1\rangle}}

\def\expect#1{{\langle#1\rangle}}

\def\tr{{\rm Tr}}
\def\H{{\hat H}}

\def\U{{\hat U}}
\def\Udag{{\hat U}^\dagger}

\def\Op{{\hat O}}

\def\Mhat{{\hat M}}
\def\Mdag{\Mhat^\dagger}
\def\Ahat{{\hat A}}
\def\Adag{\Ahat^\dagger}
\def\epsop{{\hat\varepsilon}}
\def\Xhat{{\hat X}}
\def\id{{\hat I}}

\begin{document}

\title{Infinitesimal local operations and
differential conditions for entanglement monotones}
\author{Ognyan Oreshkov}
\email{oreshkov@usc.edu}
\affiliation{Department of Physics, University of Southern California,
Los Angeles, CA  90089}
\author{Todd A. Brun}
\email{tbrun@usc.edu}
\affiliation{Communication Sciences Institute, University of Southern California,
Los Angeles, CA  90089}

\date{\today}

\begin{abstract}
Much of the theory of entanglement concerns the transformations
that are possible to a state under local operations with classical
communication (LOCC); however, this set of operations is
complicated and difficult to describe mathematically.  An idea
which has proven very useful is that of the {\it entanglement
monotone}:  a function of the state which is invariant under local
unitary transformations and always decreases (or increases) on
average after any local operation.  In this paper we look on LOCC
as the set of operations generated by {\it infinitesimal local
operations}, operations which can be performed locally and which
leave the state little changed.  We show that a necessary and
sufficient condition for a function of the state to be an
entanglement monotone under local operations that do not involve
information loss is that the function be a monotone under
infinitesimal local operations.  We then derive necessary and
sufficient differential conditions for a function of the state to
be an entanglement monotone.  We first derive two conditions for
local operations without information loss, and then show that they
can be extended to more general operations by adding the
requirement of {\it convexity}.  We then demonstrate that a number
of known entanglement monotones satisfy these differential
criteria. Finally, as an application, we use the differential
conditions to construct a new polynomial entanglement monotone for
three-qubit pure states. It is our hope that this approach will
avoid some of the difficulties in the theory of multipartite and
mixed-state entanglement.
\end{abstract}

\maketitle


\section{Introduction}

One of the most impressive achievements of quantum information
theory is the theory of entanglement.  The theory of entanglement
concerns the transformations that are possible to a state under
local operations with classical communication (LOCC).  The
paradigmatic experiment is a quantum system comprising several
subsystems, each in a separate laboratory under control of a
different experimenter:  Alice, Bob, Cara, etc.  Each experimenter
can perform any physically allowed operation on his or her
subsystem---unitary transformations, generalized measurements,
indeed any trace-preserving completely positive operation--and
communicate their results to each other without restriction.  They
are not, however, allowed to bring their subsystems together and
manipulate them jointly.  An LOCC protocol consists of any number
of local operations, interspersed with any amount of classical
communication; the choice of operations at later times may depend
on the outcomes of measurements at any earlier time.

The results of Bennett et al. \cite{Bennett96a,Bennett96b,Bennett96c} and Nielsen \cite{Nielsen99},
among many others \cite{Vidal99,Jonathan99a,Hardy99,Jonathan99b,Vidal00a}, have given us a nearly complete
theory of entanglement for {\it bipartite} systems in pure states.
Unfortunately, great difficulties have been encountered in trying
to extend these results both to mixed states and to states with
more than two subsystems ({\it multipartite} systems).  The
reasons for this are many; but one reason is that the set LOCC is
complicated and difficult to describe mathematically \cite{Bennett99}.

One mathematical tool which has proven very useful is that of the
{\it entanglement monotone}:  a function of the state which is
invariant under local unitary transformations and always decreases
(or increases) on average after any local operation.  These
functions were described by Vidal \cite{Vidal00b}, and large
classes of them have been enumerated since then.

We will consider those protocols in LOCC that preserve pure states
as the set of operations generated by {\it infinitesimal local
operations}:  operations which can be performed locally and which
leave the state little changed including infinitesimal local
unitaries and weak generalized measurements.  In Bennett et al.
\cite{Bennett99} it was shown that infinitesimal local operations
can be used to perform any local operation with the additional use
of local ancillary systems--extra systems residing in the local
laboratories, which can be coupled to the subsystems for a time
and later discarded. Recently we have shown that any local
generalized measurement can be implemented as a sequence of weak
measurements {\it without} the use of ancillas \cite{Oreshkov05}. This
implies that a necessary and sufficient condition for a function
of the state to be a monotone under local operations that preserve
pure states is the function to be a monotone under infinitesimal
local operations.

In this paper we derive differential conditions for a function of
the state to be an entanglement monotone by considering the change
of the function on average under infinitesimal local operations up
to the lowest order in the infinitesimal parameter.  We thus
obtain conditions that involve at most second derivatives of the
function.  We then prove that these conditions are both necessary
and sufficient. We show that the conditions are satisfied by a
number of known entanglement monotones and we use them to
construct a new polynomial entanglement monotone for three-qubit
pure states.

It is our hope that this approach will provide a new window with
which to study LOCC, and perhaps avoid some of the difficulties in
the theory of multipartite and mixed-state entanglement.  By
looking only at the differential behavior of entanglement
monotones, we avoid concerns about the global structure of LOCC.

In section II, we define the basic concepts of this paper:  LOCC
operations, entanglement monotones, and infinitesimal operations.
In section III, we show how all local operations that preserve
pure states can be generated by a sequence of infinitesimal local
operations. In section IV, we derive differential conditions for a
function of the state to be an entanglement monotone. There are
two such conditions for pure-state entanglement monotones: the
first guarantees invariance under local unitary transformations
(LU invariance), and involves only the first derivatives of the
function, while the second guarantees monotonicity under local
measurements, and involves second derivatives. For mixed-state
entanglement monotones we add a further condition, {\it
convexity}, which ensures that a function remains monotonic under
operations that lose information (and can therefore transform pure
states to mixed states). In section V, we look at some known
monotones--the norm of the state, the local purity, and the
entropy of entanglement--and show that they obey the differential
criteria. In section VI, we use the differential conditions to
construct a new polynomial entanglement monotone for three-qubit
pure states which depends on the invariant identified by Kempe
\cite{Kempe99}. Finally, in section VII we conclude. In the
appendix, we show that higher derivatives of the function are not
needed to prove monotonicity.

\section{Basic definitions}

\subsection{LOCC}

An operation (or protocol) in LOCC consists of a sequence of local
operations with classical communication between them.  Initially,
we will consider only those local operations that preserve pure
states: {\it unitaries}, in which the state is transformed
\begin{equation}
\rho \rightarrow \U\rho\Udag ,\ \ \Udag\U = \U\Udag= \id ,
\end{equation}
and {\it generalized measurements}, in which the state randomly changes
\begin{equation}
\rho \rightarrow \rho_j = \Mhat_j \rho \Mhat^{\dagger}_j /p_j ,\ \
\sum_j \Mdag_j\Mhat_j = \id ,
\end{equation}
with probability $p_j = \tr\left\{\Mdag_j\Mhat_j\rho\right\}$,
where the index $j$ labels the possible outcomes of the
measurement. Note that we can think of a unitary as being a
special case of a generalized measurement with only one possible
outcome. One can think of this class of operations as being
limited to those which do not discard information.  Later, we will
relax this assumption to consider general operations, which can
take pure states to mixed states. Such operations do involve loss
of information. Examples include performing a measurement without
retaining the result, performing an unknown unitary chosen at
random, or entangling the system with an ancilla which is
subsequently discarded.

The requirement that an operation be local means that the
operators $\U$ or $\Mhat_j$ must have a tensor-product structure
$\U \equiv \U\otimes\id$, $\Mhat_j \equiv \Mhat_j \otimes \id$,
where they act as the identity on all except one of the
subsystems.  The ability to use classical communication implies
that the choice of later local operations can depend arbitrarily
on the outcomes of all earlier measurements.  One can think of an
LOCC operation as consisting of a series of ``rounds.''  In each
round, a single local operation is performed by one of the local
parties; if it is a measurement, the outcome is communicated to
all parties, who then agree on the next local operation.

\subsection{Entanglement monotones}

For the purposes of this paper, we define an entanglement monotone
to be a real-valued function of the state with the following
properties:  if we start with the system in a state $\rho$ and
perform a local operation which leaves the system in one of the
states $\rho_1,\cdots,\rho_n$ with probabilities $p_1,\ldots,p_n$,
then the value of the function must not increase on average:
\begin{subequations}
\label{eq:EM}
\begin{equation}
f(\rho) \ge \sum_j p_j f(\rho_j) . \label{monotonicity1}
\end{equation}

Furthermore, we can start with a state selected randomly from an
ensemble $\{\rho_k, p_k\}$. If we dismiss the information about
which particular state we are given (which can be done locally),
the function of the resultant state must not exceed the average of
the function we would have if we keep this information:
\begin{equation}
\sum_k p_k f(\rho_k) \geq f\left( \sum_k p_k \rho_k \right) .
\label{monotonicity2}
\end{equation}
\end{subequations}

Some functions may obey a stronger form of monotonicity, in which
the function cannot increase for any outcome:
\begin{equation}
f(\rho) \ge f(\rho_j),\  \forall j ,
\end{equation}
but this is not the most common situation.  Some monotones may be
defined only for pure states, or may only be monotonic for pure
states. In the latter case, monotonicity is defined as
non-increase on average under local operations that do not involve
information loss.

\subsection{Infinitesimal operations}

We call an operation {\it infinitesimal} if all outcomes result in
only very small changes to the state.  That is, if after an
operation the system can be left in states $\rho_1,\cdots,\rho_n$,
we must have
\begin{equation}
|| \rho - \rho_j || \ll 1, \ \forall j .
\end{equation}
For a unitary, this means that
\begin{equation}
\U = \exp(i{\epsop}) \approx \id + i {\epsop} ,
\end{equation}
where $\epsop$ is a Hermitian operator with small norm,
$||{\epsop}|| \ll 1$, ${\epsop}={\epsop}^\dagger$.  For a
generalized measurement, every measurement operator $\Mhat_j$ can
be written as
\begin{equation}
\Mhat_j = q_j (\id + {\epsop}_j ) ,
\end{equation}
where $0 \le q_j \le 1$ and ${\epsop}_j$ is an operator with small
norm $||{\epsop}_j|| \ll 1$.

Such measurements are called {\it weak}.  The term {\it weak
measurement}, however, is often taken to include measurements in
which some of the outcomes change the state a great deal, but only
with very low probability.  We do not include such measurements in
what follows.  All outcomes must leave the state almost unchanged.

\section{Local operations from infinitesimal local operations}

In this section we show how any local operation that preserves
pure states can be performed as a sequence of infinitesimal local
operations. The operations that preserve pure states are unitary
transformations and generalized measurements.

\subsection{Unitary transformations}
Every local unitary operator has the representation
\begin{equation}
\U=e^{i\H},
\end{equation}
where $\H$ is a local hermitian operator. We can write
\begin{equation}
\U=\lim_{n\rightarrow\infty}(\id+i\H/n)^n,
\end{equation}
and define
\begin{equation}
\epsop=\H/n
\end{equation}
for a suitably large value of $n$.  Thus, in the limit $n\rightarrow\infty$,
any local unitary operation can be thought of as an infinite
sequence of infinitesimal local unitary operations driven by
operators of the form
\begin{equation}
\U_{\varepsilon}\approx \id+i\epsop,
\end{equation}
where $\epsop$ is a small ($\|{\epsop}\|\ll 1$) local
hermitian operator.

\subsection{Generalized measurements}

Recently it has been shown \cite{Oreshkov05} that any local measurement
can be generated by a sequence of weak local measurements. Since a
measurement with any number of outcomes can be implemented as a
sequence of two-outcome measurements, it suffices to show this for
generalized measurements with two outcomes.

If the initial state of the system has a density matrix $\rho$,
the two possible outcomes of a measurement with operators $\Mhat_1$
and $\Mhat_2$ are $\Mhat_1\rho \Mdag_1/p_1$ and $\Mhat_2\rho
\Mdag_2/p_2$, where $p_{1,2}=\tr(\Mhat_{1,2}\rho
\Mdag_{1,2})$ are the corresponding probabilities. Using the polar
decomposition, the two measurement operators can be written as
$\Mhat_1=\U_1\sqrt{\Mhat_1^{\dagger}\Mhat_1}$ and
$\Mhat_2=\U_2\sqrt{\Mhat_2^{\dagger}\Mhat_2}$, where $\U_1$ and $\U_2$ are
unitary.  Since we have already seen that we can do any local unitary transformation
by a sequence of infinitesimal steps, if we can first measure
the positive operators $\sqrt{\Mhat_1^\dagger \Mhat_1}$ and
$\sqrt{\Mhat_2^\dagger \Mhat_2}$ by a series of
infinitesimal steps, we can then apply $\U_1$ or $\U_2$ (conditional on the outcome),
and can therefore measure $\Mhat_1$ and $\Mhat_2$ by infinitesimal steps as well.
So without loss of generality, we consider only positive
measurement operators:  $\Mhat_j=\Mhat_j^\dagger$, $\id \ge \Mhat_j \ge 0$.
Note that in this case, $\Mhat_1$ and $\Mhat_2$ commute:  $\Mhat_1\Mhat_2=\Mhat_2\Mhat_1$.

We now decompose this measurement into a series of weak measurements.  We can think of
the procedure as a random walk along a curve in state space; the position on this curve is
indicated by a single parameter $x$, with $x=0$ being the initial state.
The current state of the system at any point during the procedure can be written
\begin{equation}
\Mhat(x)\rho \Mhat(x)/\tr(\Mhat^2(x)\rho),
\end{equation}
where
\begin{equation}
\Mhat(x)=\sqrt{\frac{\id+\tanh(x)(\Mhat_2^2-\Mhat_1^2)}{2}},\ \
x\in R.
\end{equation}
In the limit $x\rightarrow\pm\infty$ the effective operator reduces to $\Mhat_{1,2}$.

In \cite{Oreshkov05} it has been shown that depending on the current
value of the parameter $x$, one can perform a two-outcome
measurement on the system with positive operators
$\Mhat(x,\pm\epsilon)$ which satisfy
\begin{eqnarray}
[\Mhat(x,\epsilon)]^2 + [\Mhat(x,-\epsilon)]^2 &=& \id, \nonumber\\
\Mhat(x,\pm\epsilon)\Mhat(x) &\propto& \Mhat(x\pm\epsilon).
\label{recur}
\end{eqnarray}
Since the state is normalized, the factor of proportionality is
irrelevant; the two possible outcomes simply change the parameter
$x$ by $+\epsilon$ or $-\epsilon$. Thus the measurement procedure
is a random walk along the curve $\Mhat(x)$, with a step size
$|\epsilon|$. We continue this walk until $|x| \ge X$ for some $X$
which is sufficiently large that $\Mhat(-X) \approx \Mhat_1$ and
$\Mhat(X) \approx \Mhat_2$, to whatever precision we desire. It
has been shown that the probabilities of the outcomes for this
procedure are exactly the same as those for a single generalized
measurement. The exact form of the operators
$\Mhat(x,\pm\epsilon)$ is derived in \cite{Oreshkov05}:
\begin{equation}
\Mhat(x,\pm\epsilon) = \sqrt{ C_\pm
  \frac{\id+\tanh(x\pm\epsilon)(\Mhat_2^2-\Mhat_1^2)}{\id+\tanh(x)(\Mhat_2^2-\Mhat_1^2)}},
\label{step_operator}
\end{equation}
where the weights $C_\pm$ are chosen to ensure that these operators form a generalized measurement:
\begin{equation}
C_\pm = (1\pm\tanh(\epsilon)\tanh(x))/2  .
\end{equation}
For any finite $x$,
$\Mhat(x)^{-1}$ is well defined, so from \eqref{recur} and \eqref{step_operator} it is easy to
see that if $|\epsilon|\ll 1$, we have $\Mhat(x,\epsilon) =
\sqrt{1/2}(\id+O(\epsilon))$:  the measurements are weak. Thus
every measurement can be implemented as a sequence of weak
measurements. Moreover, if the original measurement is local, the
weak measurements are also local.

Clearly, the fact that infinitesimal local operations are part of
the set of LO means that an entanglement monotone must be a
monotone under infinitesimal local operations. The result
discussed in this section implies that if a function is a monotone
under infinitesimal local unitaries and generalized measurements,
it is a monotone under all local unitaries and generalized
measurements (the operations that do not involve information loss
and preserve pure states). Based on this result, in the next
section we derive necessary and sufficient conditions for a
function to be an entanglement monotone.

\section{Differential conditions for entanglement monotones}

Let us now consider the change in the state under an infinitesimal
local operation. Without loss of generality, we assume that the
operation is performed on Alice's subsystem. In this case, it is
convenient to write the density matrix of the system as
\begin{equation}
{\rho}=\underset{i,j,l,m}\sum \rho_{ijlm} |i_A\rangle \langle
l_A|\otimes |j_{BC...}\rangle \langle m_{BC...}|,
\end{equation}
where the set $\{|i_A\rangle\}$ and the set
$\{|j_{BC...}\rangle\}$ are arbitrary orthonormal bases for
subsystem $A$ and the rest of the system, respectively. Any
function of the state $f(\rho)$ can be thought of as a function of
the coefficients in the above decomposition:
\begin{equation}
f(\rho) = f(\rho_{ijlm}).
\end{equation}

\subsection{Local unitary invariance}

Unitary operations are invertible, and therefore the monotonicity
condition reduces to an invariance condition for LU
transformations. Under local unitary operations on subsystem $A$
the components of ${\rho}$ transform as follows:
\begin{equation}
\rho_{ijlm} \rightarrow \underset{k,p}\sum
U_{ik}\rho_{kjpm}U^*_{lp},
\end{equation}
where $U_{ik}$ are the components of the local unitary operator in
the basis $\{|i_A\rangle\}$. We consider infinitesimal local
unitary operations:
\begin{equation}
U_{lk} = \left(e^{i\epsop}\right)_{lk},
\end{equation}
where ${\epsop}$ is a local hermitian operator acting on
subsystem $A$, and
\begin{equation}
\|{\epsop}\|\ll 1.
\end{equation}
Up to first order in $\epsop$ the coefficients $\rho_{ijlm}$
transform as
\begin{equation}
\rho_{ijlm} \rightarrow \rho_{ijlm} + i[\epsop, \rho]_{ijlm}.
\end{equation}
Requiring LU-invariance of $f(\rho)$, we obtain that the function
must satisfy
\begin{equation}
\underset{i,j,l,m}\sum\frac{\partial f}{\partial
\rho_{ijlm}}[\epsop, \rho]_{ijlm}=0. \label{three}
\end{equation}
Analogous equations must be satisfied for arbitrary hermitian
operators $\epsop$ acting on the other parties' subsystems. In a
more compact form, the condition can be written as
\begin{equation}
\tr\left\{ \frac{\partial f}{\partial\rho}[\epsop , \rho] \right\}
= 0, \label{mixed1}
\end{equation}
where $\epsop$ is an arbitrary local hermitian operator.

\subsection{Non-increase under infinitesimal local measurements}

As mentioned earlier, a measurement with any number of outcomes
can be implemented as a sequence of measurements with two
outcomes, and a general measurement can be done as a measurement
with positive operators, followed by a unitary conditioned on the
outcome; therefore, it suffices to impose the monotonicity
condition for two-outcome measurements with positive measurement
operators. Consider local measurements on subsystem $A$ with two
measurement outcomes, given by operators ${\Mhat}_1^2+{\Mhat}_2^2
= \id$. Without loss of generality, we assume
\begin{eqnarray}
{\Mhat}_1 &=&
\sqrt{ ({\id}+{\epsop})/2 },\nonumber\\
{\Mhat}_2 &=&
\sqrt{ ({\id}-{\epsop})/2 },
\label{mm}
\end{eqnarray}
where ${\epsop}$ is again a small local hermitian operator acting
on $A$ (in the previous section we saw that any two-outcome
measurement with positive operators can be generated by weak
measurements of this type). Upon measurement, the state undergoes
one of two possible transformations
\begin{eqnarray}
\rho &\rightarrow& \frac{\Mhat_{1,2}\rho \Mhat_{1,2}}{p_{1,2}},
\end{eqnarray}
with probabilities $ p_{1,2} = \tr\left\{ {\Mhat_{1,2}}^2 \rho
\right\}$. Since $\epsop$ is small, we can expand
\begin{eqnarray}
{\Mhat}_1 &=& \frac{1}{\sqrt{2}}({\id} + {\epsop}/2 - {\epsop}^2/8 - \cdots), \label{m1} \\
{\Mhat}_2 &=&\frac{1}{\sqrt{2}}({\id} - {\epsop}/2 - {\epsop}^2/8 - \cdots).\label{m2}
\end{eqnarray}
The condition for non-increase on average of the function $f$
under infinitesimal local measurements is
\begin{equation}
p_1 f(\Mhat_1\rho \Mhat_1/p_1) + p_2 f(\Mhat_2\rho \Mhat_2/p_2)
\le f(\rho) . \label{nonincrease4}
\end{equation}
Expanding (\ref{nonincrease4}) in powers of $\epsop$ up to second
order, we obtain
\begin{equation}
\frac{1}{4}\tr\left\{ \frac{\partial f}{\partial\rho}[[\epsop,
\rho],\epsop] \right\}
  + \tr\left\{ \frac{\partial^2 f}{\partial\rho^{\otimes2}}
  \left( \tr(\epsop\rho)\rho - \frac{1}{2}\{\epsop,\rho\} \right)^{\otimes 2} \right\}
  \leq 0 ,
\label{mixed3}
\end{equation}
where $\{\epsop, \rho \}$ is the anti-commutator of $\epsop$ and
$\rho$. The inequality must be satisfied for an arbitrary local
hermitian operator $\epsop$.

So long as (\ref{mixed3}) is satisfied by a strict inequality, it
is obvious that we need not consider higher-order terms in
$\epsop$. But what about the case when the condition is satisfied
by equality?  In the Appendix we will show that even in the case
of equality, (\ref{mixed3}) is still the necessary and sufficient
condition for monotonicity under local generalized measurements.
There we also prove the sufficiency of the LU-invariance condition
\eqref{mixed1}. This allows us to state the following
\\\\
\textbf{Theorem}: A twice-differentiable function $f(\rho)$ of the
density matrix is a monotone under local unitary operations and
generalized measurements, if and only if it satisfies
\eqref{mixed1} and \eqref{mixed3}.
\\\\
Unitary operations and generalized measurements are the operations
that preserve pure states. Other operations (which involve loss of
information), such as positive maps, would in general cause pure
states to evolve into mixed states. A measure of pure-state
entanglement need not be defined over the entire set of density
matrices, but only over pure states. Thus a measure of pure-state
entanglement, when expressed as a function of the density matrix,
may have a significantly simpler form than its generalizations to
mixed states. For example, the entropy of entanglement for
bipartite pure states can be written in the well-known form
$S_A(\rho)=-\tr(\rho_A\log \rho_A)$, where $\rho_A$ is the reduced
density matrix of one of the parties' subsystems. When directly
extended over mixed states, this function is not well justified,
since $S_A(\rho)$ may have a different value from $S_B(\rho)$.
Moreover, $S_A(\rho)$ by itself is not a mixed-state entanglement
monotone, since it may increase under local positive maps on
subsystem A (these properties of the entropy of entanglement will
be discussed further in section V). One generalization of the
entropy of entanglement to mixed states is the entanglement of
formation \cite{Bennett96c}, which is defined as the minimum of
$\sum_i p_i S_A(\rho_i)$ over all ensembles of bipartite pure
states $\{\rho_i, p_i\}$ realizing the mixed state: $\rho=\sum_i
p_i \rho_i$. This quantity is a mixed-state entanglement monotone.
As a function of $\rho$, it has a much more complicated form than
the above expression for the entropy of entanglement. In fact,
there is no known analytic expression for the entanglement of
formation in general. The problem of extending pure-state
entanglement monotones to mixed states is an important one, since
every mixed-state entanglement monotone can be thought of as an
extension of a pure-state entanglement monotone. Note, however,
that a pure-state entanglement monotone may have many different
mixed-state generalizations. The relation between the entanglement
of formation and the entropy of entanglement presents one way to
perform such an extension (convex-roof extension). For every
pure-state entanglement monotone $m(\rho)$, one can define a
mixed-state extension $M(\rho)$ as the minimum of $\sum_i p_i
m(\rho_i)$ over all ensembles of pure states $\{\rho_i, p_i\}$
realizing the mixed state: $\rho=\sum_i p_i \rho_i$. It is easy to
verify that $M(\rho)$ is an entanglement monotone for mixed
states. On the set of pure states the function $M(\rho)$ reduces
to $m(\rho)$. As the example with the entropy of entanglement
suggests, not every form of a pure-state entanglement monotone
corresponds to a mixed-state entanglement monotone when trivially
extended to all states -- there are additional conditions that a
mixed-state entanglement monotone must satisfy. On the basis of
the above considerations, it makes sense to consider separate sets
of differential conditions for pure-state and mixed-state
entanglement monotones.
\\\\
\textbf{Corollary 1:} A twice-differentiable function $f(\rho)$ of
the density matrix is a pure-state entanglement monotone, if and
only if it satisfies \eqref{mixed1} and \eqref{mixed3} for pure
$\rho$.
\\\\
For pure states $\rho = \ket\psi\bra\psi$, the elements of $\rho$
are $\rho_{ij\ell m} = \alpha_{ij}\alpha^*_{\ell m}$, where the
$\{\alpha_{ij}\}$ are the state amplitudes: $ |\psi\rangle =
\underset{i,j}{\sum}\alpha_{ij}|i_A\rangle |j_{BC...}\rangle $.
Any function on pure states $f(\rho)\equiv f(|\psi\rangle)$ is
therefore a function of the state amplitudes and their complex
conjugates:
\begin{equation}
f(|\psi\rangle) = f(\{\alpha_{ij}\}, \{\alpha^{\ast}_{ij}\}).
\end{equation}
By making the substitution $\rho_{ij\ell m} =
\alpha_{ij}\alpha^*_{\ell m}$ into (\ref{mixed1}) and
(\ref{mixed3}), we can (after considerable algebra) derive
alternative forms of the differential conditions for functions of
the state vector:
\begin{equation}
\underset{i,j,k}\sum\frac{\partial f}{\partial \alpha_{ij}}
\varepsilon_{ik} \alpha_{kj}=\underset{i,j,k}\sum\frac{\partial
f}{\partial
\alpha^{\ast}_{ij}}\varepsilon^{\ast}_{ik}\alpha^{\ast}_{kj},
\label{LU}
\end{equation}
\begin{equation}
\underset{i,j,k,l,m,n}\sum\frac{\partial^2f}{\partial
\alpha_{ij}\partial\alpha_{mn}}
  \left( \varepsilon_{ik}\alpha_{kj}-\expect{\epsop}\alpha_{ij} \right)
  \left( \varepsilon_{m\ell}\alpha_{\ell n}-\expect{\epsop}\alpha_{mn} \right)
+ c.c. \leq 0. \label{nonincrease3}
\end{equation}
Here $\epsop$ is a local hermitian operator acting on subsystem A.
Analogous conditions must be satisfied for $\epsop$ acting on the
other parties' subsystems.

\subsection{Monotonicity under operations with information
loss}

Besides monotonicity under local unitaries and generalized
measurements, an entanglement monotone for mixed states should
also satisfy monotonicity under local operations which involve
{\it loss of information}. The most general transformation that
involves loss of information has the form
\begin{equation}
\rho \rightarrow \rho_k = \frac{1}{p_k} \sum_j \Ahat_{k,j} \rho \Adag_{k,j} ,
\end{equation}
where
\begin{equation}
p_k = \tr\left\{ \sum_j \Ahat_{k,j} \rho \Adag_{k,j} \right\}
\end{equation}
is the probability for outcome $k$.  The operators $\{\Ahat_{k,j}\}$ must satisfy
\begin{equation}
\sum_{k,j} \Adag_{k,j} \Ahat_{k,j} = \id .
\end{equation}
We can see that this includes unitary transformations, generalized
measurements, and completely positive trace-preserving maps as
special cases.

It occasionally makes sense to consider even more general
transformations, where the operators need not sum to the identity:
\begin{equation}
\sum_{k,j} \Adag_{k,j} \Ahat_{k,j} \le \id .
\end{equation}
This corresponds to a situation where only certain outcomes are
retained, and others are discarded; the probabilities add up to
less than 1 due to these discarded outcomes.  We say such a
transformation involves {\it postselection}.

With or without postselection, we are concerned with the case
where all operations are done locally, so that all the operators
$\{\Ahat_{k,j}\}$ act on a single subsystem. Every such
transformation can be implemented as a sequence of local
generalized measurements (possibly discarding some of the
outcomes) and local completely positive maps. In operator-sum
representation, a completely positive map can be written
\begin{equation}
\rho \rightarrow \sum_k \Mhat_k \rho \Mhat_k^{\dagger},
\end{equation}
where
\begin{equation}
\sum_k \Mhat_k^{\dagger}\Mhat_k \leq \id.
\label{positivity}
\end{equation}
Therefore, in addition to \eqref{mixed1} and \eqref{mixed3} we must impose the condition
\begin{equation}
f(\rho) \geq f\left( \sum_k \Mhat_k\rho \Mhat_k^{\dagger} \right) .
\label{maps}
\end{equation}
for all sets of local operators $\{\Mhat_k\}$ satisfying
(\ref{positivity}).

Suppose the parties are supplied with a state $\rho_k$ taken from an ensemble
$\{\rho_k, p_k\}$.  Discarding the information of the actual state amounts to the transformation
\begin{equation}
\{\rho_k, p_k\} \rightarrow \rho '=\underset{k}{\sum} p_k \rho_k.
\end{equation}
As pointed out in \cite{Vidal00b}, discarding information should not increase
the entanglement of the system on average.
Therefore, for any ensemble $\{\rho_k, p_k\}$, an entanglement
monotone on mixed states should be {\it convex}:
\begin{equation}
\sum_k p_k f(\rho_k) \geq f\left( \sum_k p_k \rho_k \right) .
\label{convex}
\end{equation}
Condition \eqref{convex}, together with condition (\ref{mixed3})
for monotonicity under local generalized measurements, implies
monotonicity under local completely positive maps:
\begin{equation}
f\left( \sum_k \Mhat_k\rho \Mhat_k^{\dagger} \right)
  \leq \sum_k p_k f \left( \frac{\Mhat_k\rho \Mhat^{\dagger}_k}{p_k} \right)
  \leq f(\rho).
\end{equation}
It is easy to see that if this inequality holds without postselection,
it must also hold with postselection.

It follows that a function of the density matrix is an
entanglement monotone for mixed states if and only if it is (1) a
convex function on the set of density matrices and (2) a monotone
under local unitaries and generalized measurements. Fortunately,
there are also simple differential conditions for convexity.  A necessary
and sufficient condition for a twice-differentiable function of
multiple variables to be convex on a convex set is that its Hessian
matrix be positive on the interior of the convex set (in this
case, the set of density matrices). Therefore, in addition to
\eqref{mixed1} and \eqref{mixed3} we add the differential condition
\begin{equation}
\tr\left\{ \frac{\partial^2 f(\rho)}{\partial\rho^{\otimes
2}}\sigma^{\otimes 2} \right\} \geq 0,
\label{mixed4}
\end{equation}
which must be satisfied at every $\rho$ on the interior of the set
of density matrices for an arbitrary traceless hermitian matrix
$\sigma$.
\\\\
\textbf{Corollary 2:} A twice-differentiable function $f(\rho)$ of
the density matrix is a mixed-state entanglement monotone, if and
only if it satisfies \eqref{mixed1}, \eqref{mixed3} and
\eqref{mixed4}.

\section{Examples}

In this section we demonstrate how conditions \eqref{mixed1},
\eqref{mixed3} and \eqref{mixed4} can be used to verify if a
function is an entanglement monotone. We show this for three well
known entanglement monotones: the norm of the state of the system,
the trace of the square of the reduced density matrix of any
subsystem, and the entropy of entanglement. In the next section we
will use some of the observations made here to construct a new
polynomial entanglement monotone for three-qubit pure states.

\subsection{Norm of the state}

The most trivial example is the norm or the trace of the density
matrix of the system:
\begin{equation}
I_1=\tr\{\rho\}.
\end{equation}
Clearly $I_1$ is a monotone under LOCC, since all operations that we
consider either preserve or decrease the trace. But for the
purpose of demonstration, let us verify that $I_1$ satisfies the
differential conditions.

The LU-invariance condition \eqref{mixed1} reads
\begin{equation}
\tr\left\{ \frac{\partial I_1}{\partial\rho}[\epsop , \rho] \right\}
  = \tr\left\{ [\epsop , \rho ] \right\} = 0.
\end{equation}
The second equality follows from the cyclic invariance of the
trace.

Since the trace is linear, the second term in condition
\eqref{mixed3} vanishes, and we consider only the first term:
\begin{equation}
\tr\left\{ \frac{\partial I_1}{\partial\rho}[[\epsop, \rho],\epsop] \right\}
  = \tr\left\{ [[\epsop, \rho],\epsop] \right\} = 0.
\end{equation}
The condition is satisfied with equality, again due to the cyclic
invariance of the trace, implying that the norm remains invariant
under local measurements. The convexity condition \eqref{mixed4}
is also satisfied by equality.

\subsection{Local purity}

The second example is the purity of the reduced density matrix:
\begin{equation}
I_2=\tr\left\{ \rho_A^2 \right\},
\end{equation}
where $\rho_A$ is the reduced density matrix of subsystem $A$
(which in general need not be a one-party subsystem).  Note that
this is an {\it increasing} entanglement monotone for pure
states---the purity of the local reduced density matrix can only
increase under LOCC.

It has been shown in \cite{Brun04} that every $m$-th degree
polynomial of the components of the density matrix $\rho$ can be
written as an expectation value of an observable $\Op$ on $m$
copies of $\rho$:
\begin{equation}
f(\rho)=\tr\left\{ \Op \rho^{\otimes m} \right\} .
\end{equation}
Here we have
\begin{equation}
\tr\left\{ \rho_A^2 \right\}
  = \tr\left\{ \hat{C} \rho^{\otimes 2} \right\} ,
\end{equation}
where the components of $\hat{C}$ are
\begin{equation}
C_{lpsnkjqm}=\delta_{jp}\delta_{mn}\delta_{lq}\delta_{ks}.
\end{equation}
Therefore
\begin{eqnarray}
\tr\left\{ \frac{\partial I_2}{\partial\rho}[\epsop , \rho] \right\}
  &=&  \tr\left\{ \hat{C}\left( [\epsop , \rho ]\otimes\rho
    + \rho\otimes[\epsop , \rho ] \right) \right\} \nonumber\\
  &=& \tr_A \left\{ [\epsop , \rho ]_A \rho_A + \rho_A [\epsop,\rho]_A \right\} \nonumber\\
  &=& 2 \tr_A \left\{ \rho_A [\epsop , \rho]_A \right\} ,
\label{ifour}
\end{eqnarray}
where by $\Op_A$ we denote the partial trace of an operator $\Op$ over all subsystems except $A$.
If $\epsop$ does not act on subsystem $A$, then
$[\epsop,\rho ]_A = 0$ and the above expression vanishes. If it acts on
subsystem $A$, then $[\epsop , \rho ]_A =[\epsop,\rho_A]$
and the expression vanishes due to the cyclic invariance of the trace.

Now consider condition \eqref{mixed3}. If $\epsop$ does not act on
subsystem $A$, then
\begin{equation}
[[\epsop, \rho],\epsop]_A=0.\label{doublecom}
\end{equation}
From \eqref{mixed3} we get
\begin{eqnarray}
0 &\le& \frac{1}{4}\tr\left\{ \frac{\partial I_2}{\partial\rho}
[[\epsop, \rho],\epsop] \right\}
  + \tr\left\{ \frac{\partial^2 I_2}{\partial\rho^{\otimes2}} \left( \tr\left\{ \epsop\rho \right\} \rho
  - \frac{1}{2} \{\epsop,\rho\} \right)^{\otimes 2} \right\} \nonumber\\
&& = 2\tr\left\{ \left( \tr\{\epsop\rho\} \rho
  - \frac{1}{2}\{\epsop,\rho\} \right)_A^2 \right\}.
\end{eqnarray}
The inequality follows from the fact that
$\left(\tr\{\epsop\rho\}\rho - (1/2) \{\epsop,\rho\}\right)_A^2$
is a positive operator.

If $\epsop$ acts on $A$, we can use the fact that for pure states
\begin{equation}
\tr\left\{ \rho_A^2 \right\}= \tr\left\{ \rho_B^2 \right\},
\end{equation}
where $B$ denotes the subsystem complementary to $A$. Then we can
apply the same argument as before for the function $\tr\left\{
\rho_B^2 \right\}$. Therefore $I_2$ does not \emph{decrease} on
average under local generalized measurements, and is an
entanglement monotone for pure states.

What about mixed states?  For \emph{increasing} entanglement monotones
the convexity condition \eqref{mixed4} becomes a \emph{concavity} condition---the
direction of the inequality is inverted. In the case of $I_2$, however, we have
\begin{equation}
\tr\left\{ \frac{\partial^2 I_2(\rho)}{\partial\rho^{\otimes
2}}\sigma^{\otimes 2} \right\}
  = 2\tr\left\{ \sigma_A^2 \right\} \geq 0,
\end{equation}
i.e., the function is convex. This means that
$\tr\{\rho_A^2\}$ is \emph{not} a good measure of entanglement for
mixed states. Indeed, when extended to mixed states, $I_2$
cannot distinguish between entanglement and classical disorder.

\subsection{Entropy of entanglement}

Finally consider the von~Neumann entropy of entanglement:
\begin{equation}
S_A=-\tr(\rho_A\log \rho_A).
\end{equation}
Expanding around $\rho_A=\id$, we get
\begin{equation}
S_A=-\tr[(\rho_A-\id)+\frac{1}{2}(\rho_A-\id)^2-\frac{1}{6}(\rho_A-\id)^3
+ ...].
\end{equation}
The LU-invariance follows from the fact that every term in this
expansion satisfies \eqref{mixed1}. If we substitute the $n$-th
term in the condition, we obtain
\begin{equation}
\tr([\epsop,\rho]_A (\rho_A-\id)^{n-1})=0.
\end{equation}
This is true either because $[\epsop,\rho]_A=0$ when
$\epsop$ does not act on $A$, or because otherwise
$[\epsop,\rho]_A= [\epsop,\rho_A]$ and the equation
follows from the cyclic invariance of the trace.

Now to prove that $S_A$ satisfies \eqref{mixed3}, we will first
assume that $\rho_A^{-1}$ exists. Then we can formally write
\begin{equation}
\frac{\partial}{\partial \rho}\log{\rho_A} = \frac{\partial\rho_A}{\partial \rho}
\frac{\partial}{\partial\rho_A}\log{\rho_A} =
\frac{\partial \rho_A}{\partial\rho}\rho_A^{-1}.
\end{equation}
Consider the case when $\epsop$ does not act on $A$.
Substituting $S_A$ in \eqref{mixed3}, we get
\begin{gather}
\frac{1}{4}\tr\left\{ \frac{\partial S_A}{\partial\rho}[[\epsop,
\rho],\epsop] \right\}
  + \tr\left\{ \frac{\partial^2 S_A}{\partial\rho^{\otimes2}}
  \left( \tr\{\epsop\rho\}\rho - \frac{1}{2}\{\epsop,\rho\}\right)^{\otimes 2} \right\} \nonumber\\
= 0 + \tr\left\{ \left( \frac{\partial}{\partial\rho}\otimes \left( -\log{\rho_A}\frac{\partial \rho_A}{\partial\rho}
  -\frac{\partial \rho_A}{\partial \rho} \right) \right)
\left( \tr\{\epsop\rho\}\rho - \frac{1}{2}\{\epsop,\rho\} \right)^{\otimes 2} \right\} \nonumber\\
= -\tr\left\{ \left( \rho_A^{-1}\frac{\partial \rho_A}{\partial \rho}\frac{\partial \rho_A}{\partial \rho} \right)
  \left( \tr\{\epsop\rho\}\rho - \frac{1}{2}\{\epsop,\rho\} \right)^{\otimes 2} \right\} \nonumber\\
= - \tr_A \left\{ \rho_A^{-1}\left(  \tr\{\epsop\rho\}\rho - \frac{1}{2}\{\epsop,\rho\} \right)_A
  \left( \tr\{\epsop\rho\}\rho - \frac{1}{2}\{\epsop,\rho\} \right)_A \right\} \nonumber\\
= - \tr_A \Biggl\{ \left|\rho_A^{-1/2}\left(\tr\{\epsop\rho\}\rho -
\frac{1}{2}\{\epsop,\rho\} \right)_A\right|^2 \Biggr\} \leq 0.
\label{ent}
\end{gather}
If $\rho_A^{-1}$ does not exist, it is only on a subset of measure
zero -- where one or more of the eigenvalues of $\rho_A$ vanish.
Therefore, we can always find an arbitrarily close vicinity in the
parameters describing $\rho_A$, where $\rho_A^{-1}$ is regular and
where \eqref{mixed3} is satisfied. Since the condition is
continuous, it cannot be violated on this special subset.

If $\epsop$ acts on $A$, we can use an equivalent definition of the entropy of entanglement:
\begin{equation}
S_A=S_B=-\tr\{\rho_B\log \rho_B\},
\end{equation}
and apply the same arguments. Therefore $S_A$ is an entanglement
monotone for pure states.

The convexity condition is not satisfied, since
\begin{equation}
\tr\left\{ \frac{\partial^2 S_A}{\partial\rho^{\otimes2}}\sigma^{\otimes 2} \right\}
  = - \tr\{\rho_A^{-1}\sigma_A^2 \} \leq 0.
\end{equation}
This reflects the fact that the entropy of entanglement, like $I_2$, does not
distinguish between entanglement and classical randomness.

\section{A new entanglement monotone}

It has been shown \cite{Gingrich02} that the set of all
entanglement monotones for a multipartite pure state uniquely
determine the orbit of the state under the action of the group of
local unitary transformations. For three-qubit pure states the
orbit is uniquely determined by 5 independent continuous
invariants (not counting the norm) and one discrete invariant
\cite{Acin00,Carteret00}. Therefore, for pure states of three
qubits there must exist five independent continuous entanglement
monotones that are functions of the five independent continuous
invariants.

Any polynomial invariant in the amplitudes of a state
\[
|\psi\rangle = \underset{i,j,k\ldots}{\sum}\alpha_{ijk\ldots}|i_A\rangle
|j_B\rangle |k_C\rangle \cdots
\]
is a sum of homogenous polynomials of the form \cite{Sudbery01}
\begin{equation}
P_{\sigma\tau\cdots}(|\psi\rangle)= \alpha_{i_1 j_1 k_1 \ldots}
\alpha^*_{i_1 j_{\sigma(1)} k_{\tau(1)} \ldots} \cdots
\alpha_{i_n j_n k_n \ldots} \alpha^*_{i_n j_{\sigma(n)} k_{\tau(n)} \ldots},
\label{inv}
\end{equation}
where $\sigma, \tau, \ldots$ are permutations of (1,2,\ldots,n),
and repeated indices indicate summation. A set of five independent
polynomial invariants for three-qubit pure states is
\cite{Sudbery01}
\begin{eqnarray}
I_1&=&P_{e,(12)}\\
I_2&=&P_{(12),e}\\
I_3&=&P_{(12),(12)}\\
I_4&=&P_{(123),(132)}\\
I_5&=&| \alpha_{i_1j_1k_1} \alpha_{i_2j_2k_2} \alpha_{i_3j_3k_3}
\alpha_{i_4j_4k_4} \epsilon_{i_1i_2} \epsilon_{i_3i_4}
\epsilon_{j_1j_2} \epsilon_{j_3j_4} \epsilon_{k_1k_3}
\epsilon_{k_2k_4} |^2.
\end{eqnarray}
In the last expression $\epsilon_{ij}$ is the antisymmetric tensor
in 2 dimensions. The first three invariants are the local purities
of subsystems C, B and A, $I_4$ is the invariant identified by
Kempe \cite{Kempe99} and $I_5$ is (up to a factor) the square of
the 3-tangle identified by Coffman, Kundu and Wootters
\cite{Coffman00}. According to \cite{Gingrich02} the four known
independent continuous entanglement monotones that do not require
maximization over a multi-dimensional space are
\begin{gather}
\tau_{(AB)C}=2(1-I_1)\\
\tau_{(AC)B}=2(1-I_2)\\
\tau_{(BC)A}=2(1-I_3)\\
\tau_{ABC}= 2\sqrt{I_5},
\end{gather}
and any fifth independent entanglement monotone must depend on
$I_4$. Numerical evidence suggested that the
tenth order polynomial $\sigma_{ABC}= 3-(I_1+I_2+I_3)I_4$ might be
such an entanglement monotone. However, no rigorous proof of
monotonicity was given. Here, we will use conditions
\eqref{mixed1} and \eqref{mixed3} to construct a different independent
entanglement monotone, which is of sixth order in the amplitudes
of the state and their complex conjugates.

Observe that in \eqref{inv} the amplitudes have been combined in
such a way that subsystem A is manifestly traced out. By
appropriate rearrangement, one can write the same expression in a
form where an arbitrary subsystem is manifestly traced out.
Therefore, any polynomial invariant can be written entirely in
terms of the components of $\tr_A\left\{\rho\right\}$ or
$\tr_B\left\{\rho\right\}$, etc. This immediately implies that the
LU-invariance condition \eqref{mixed1} is satisfied, since if
$\epsop$ acts on subsystem A, we can consider the expression in
terms of $\rho_{BC...}$, which, when substituted in
\eqref{mixed1}, would yield zero because
$[\epsop,\rho]_{BC...}=0$. It also implies that in order to prove
monotonicity under local measurements we can only consider the
second term in \eqref{mixed3}, since when $\epsop$ acts on
subsystem A, we can again consider the expression for the function
only in terms of $\rho_{BC...}$ and the first term would vanish
according to \eqref{doublecom}.

We will aim at constructing a polynomial function of three-qubit
pure states $\rho$ which has the same form when expressed in terms
of $\rho_{AB}$, $\rho_{AC}$, or $\rho_{BC}$, in order to avoid the
necessity for separate proofs of monotonicity under measurements
on the different subsystems. It has been shown in \cite{Sudbery01}
that
\begin{eqnarray}
I_4&=& 3\tr\left\{\rho_{AB}(\rho_A\otimes \rho_B)\right\} -
\tr\left\{\rho_A^3\right\} - \tr\left\{\rho_B^3\right\}\nonumber\\
&=&3\tr\left\{\rho_{AC}(\rho_A\otimes \rho_C)\right\} -
\tr\left\{\rho_A^3\right\} - \tr\left\{\rho_C^3\right\}\nonumber\\
&=&3\tr\left\{\rho_{BC}(\rho_B\otimes \rho_C)\right\} -
\tr\left\{\rho_B^3\right\} -
\tr\left\{\rho_C^3\right\}.\label{sudbery}
\end{eqnarray}
For local measurements on subsystem C it is convenient to use the first of the above expressions for $I_4$. The terms $\tr\left\{\rho_A^3\right\}$ and $\tr\left\{\rho_B^3\right\}$ are entanglement monotones by themselves. This can be easily seen by plugging them in condition \eqref{mixed3}:
\begin{gather}
\frac{1}{4}\tr\left\{ \frac{\partial
\tr\left\{\rho_{A,B}^3\right\}}{\partial\rho}[[\epsop,
\rho],\epsop] \right\}
  + \tr\left\{ \frac{\partial^2 \tr\left\{\rho_{A,B}^3\right\}}{\partial\rho^{\otimes2}}
  \left( \tr(\epsop\rho)\rho - \frac{1}{2}\{\epsop,\rho\} \right)^{\otimes 2} \right\}
\nonumber\\
  =0+6 \tr\left\{\rho_{A,B} \left( \tr\{\epsop\rho\} \rho
  - \frac{1}{2}\{\epsop,\rho\} \right)_{A,B}^2 \right\}\geq 0.
\end{gather}

These terms, however, are not independent of the invariants $I_2$
and $I_3$. The term which is independent of the other polynomial
invariants is $\tr\left\{\rho_{AB}(\rho_A\otimes\rho_B)\right\}$.
When we plug this term into condition \eqref{mixed3} we obtain an
expression which is not manifestly positive or negative. Is it
possible to construct a function dependent on this term, which
similarly to $\tr\left\{\rho_{A,B}^3\right\}$ would yield a trace
of a manifestly positive operator when substituted in
\eqref{mixed3}?

It is easy to see that if the function has the form
$\tr\left\{\Xhat^3\right\}$, where the operator $\Xhat(\rho_{AB})$
is a positive operator linearly dependent on $\rho_{AB}$, it will
be an increasing monotone under local measurements on C (for
simplicity we assume $\Xhat(0)=\hat{0}$):
\begin{gather}
\frac{1}{4}\tr\left\{ \frac{\partial\tr\left\{ \Xhat^3(\rho_{AB})
\right\}}{\partial\rho}[[\epsop,\rho],\epsop] \right\}
  + \tr\left\{ \frac{\partial^2 \tr\left\{ \Xhat^3(\rho_{AB}) \right\}}{\partial\rho^{\otimes2}}
  \left( \tr(\epsop\rho)\rho - \frac{1}{2}\{\epsop,\rho\} \right)^{\otimes 2} \right\} \nonumber\\
= 0 + 6 \tr\left\{\Xhat(\rho_{AB}) \Xhat^2(( \tr\{\epsop\rho\} \rho - \frac{1}{2}\{\epsop,\rho\} )_{AB}) \right\} \geq 0.
\end{gather}
Since we want the function to depend on $\tr\left\{\rho_{AB}(\rho_A\otimes \rho_B)\right\}$, we choose $\Xhat(\rho_{AB}) = 2\rho_{AB} + \rho_A\otimes I_B + I_A\otimes \rho_B$.  This is clearly positive for positive $\rho_{AB}$.  Expanding the trace, we obtain:
\begin{eqnarray}
\tr\left\{ \Xhat^3(\rho_{AB}) \right\} &=& 12\tr\left\{\rho_{AB}(\rho_A\otimes\rho_B) \right\}
  +12\tr\left\{ \rho_{AB}^2(I_A\otimes\rho_B)\right\}
  + 12\tr\left\{ \rho_{AB}^2(\rho_A\otimes I_B) \right\} \nonumber\\
&&  + 6\tr\left\{\rho_{AB}(I_A\otimes\rho_B)^2 \right\}
  + 6\tr\left\{\rho_{AB}(\rho_A\otimes I_B)^2 \right\}
  + 3\tr\left\{\rho_A\otimes\rho_B^2\right\} \nonumber\\
&&  + 3\tr\left\{ \rho_A^2\otimes\rho_B \right\}
  + \tr\left\{ I_A\otimes\rho_B^3 \right\}
  + \tr\left\{ \rho_A^3\otimes I_B \right\}
  + 8\tr\left\{\rho_{AB}^3 \right\} \nonumber\\
&=& 12\tr\left\{\rho_{AB}(\rho_A\otimes\rho_B) \right\}
  + 12\tr\left\{ \rho_{AB}^2(I_A\otimes\rho_B) \right\}
  + 12\tr\left\{\rho_{AB}^2(\rho_A\otimes I_B)\right\} \nonumber\\
&& +8\tr\left\{\rho_A^3\right\}
  + 8\tr\left\{\rho_B^3\right\}+8\tr\left\{\rho_{AB}^3 \right\}
  + 3\tr\left\{\rho_A^2\right\}+3\tr\left\{\rho_B^2\right\} .
\end{eqnarray}

One can show that $\tr\left\{ \rho_{AB}^2(I_A\otimes\rho_B) \right\} = \tr\left\{ \rho_{BC}(\rho_B\otimes\rho_C) \right\}$ and $\tr\left\{ \rho_{AB}^2(\rho_A\otimes I_B) \right\} = \tr\left\{ \rho_{AC}(\rho_A\otimes\rho_C) \right\}$.  We also have that $\tr\left\{ \rho_{AB}^3 \right\} = \tr\left\{
\rho_C^3 \right\}$. Using this and \eqref{sudbery}, we obtain
\begin{equation}
\tr\left\{ \Xhat^3(\rho_{AB}) \right\} = 12 I_4 + 16 \left(\tr\left\{ \rho_A^3 \right\} + \tr\left\{ \rho_B^3 \right\}
  +\tr\left\{ \rho_C^3\right\} \right) +3\tr\left\{ \rho_A^2 \right\} + 3\tr\left\{ \rho_B^2 \right\} .
\end{equation}
This expression is an increasing monotone under local measurements
on C. If we add to it $3\tr\left\{ \rho_{AB}^2\right\} =
3\tr\left\{ \rho_C^2 \right\}$, it becomes invariant under
permutations of the subsystems. Since $\tr\left\{ \rho_C^2
\right\}$ is an increasing entanglement monotone, the whole
expression will be a monotone under operations on any subsystem.
We can define the closely related quantity
\begin{equation}
\phi_{ABC}=69-\tr\left\{(2\rho_{AB} + \rho_A\otimes I_B +
I_A\otimes \rho_B)^3\right\}-3\tr\left\{ \rho_{AB}^2\right\}.
\end{equation}
This is a {\it decreasing} entanglement monotone that vanishes for product states, which is more standard for a measure of entanglement.  It depends on the invariant identified by Kempe and is therefore independent of the other known monotones for three-qubit pure states.

\section{Conclusions}

We have derived differential conditions for a twice-differentiable
function on quantum states to be an entanglement monotone.  There
are two such conditions for pure-state entanglement
monotones---invariance under local unitaries and diminishing under
local measurements---plus a third condition (overall convexity of
the function) for mixed-state entanglement monotones. We have
shown that these conditions are both necessary and sufficient.  We
then verified that the conditions are satisfied by a number of
known entanglement monotones and we used them to construct a new
polynomial entanglement monotone for three-qubit pure states.

It is our hope that this approach to the study of entanglement may
circumvent some of the difficulties that arise due the
mathematically complicated nature of LOCC.  It may be possible to
find new classes of entanglement monotones, for both pure and
mixed states, and to look for functions with particularly
desirable properties (such as additivity).  There may also be
other areas of quantum information theory where it will prove
advantageous to consider general quantum operations as continuous
processes.  This seems a very promising new direction for
research.

\section*{Acknowledgments}

The authors would like to thank Stephen Adler, Dorit Aharonov,
Lane Hughston, R\"udiger Schack, Martin Varbanov and Guifr\'e
Vidal for useful conversations.  This research was supported in
part by the Martin A. and Helen Chooljian Membership in Natural
Sciences at the Institute for Advanced Study, a USC Zumberge Grant
for Innovation, and NSF Grant No.~EMT-0524822.

After this paper was written, we became aware of the paper
\cite{Plenio05} by Plenio, which claims the existence of
non-convex entanglement monotones. We would like to point out that
such a conclusion arises from a different definition of
entanglement monotones - namely, functions that obey
\eqref{monotonicity1} but not necessarily \eqref{monotonicity2}.
At present, we do not know of any argument against imposing
\eqref{monotonicity2}, which was originally given in
\cite{Vidal00b} and is briefly discussed in this paper. Since
ultimately the requirement of \eqref{monotonicity2} is a matter of
definition, just like condition \eqref{monotonicity1}, we have
conformed to the definition of entanglement monotones given in
\cite{Vidal00b}.

In this version of the paper we have corrected three minor
mistakes that appear in the published article. The first one was a
missing factor of 2 in condition \eqref{mixed3}, the second one
was in the differential form of the convexity condition
\eqref{mixed4}, and the third one was in the proof that the local
purity is an entanglement monotone for pure states. Neither change
alters the conclusions of the paper.

\section*{Appendix:  Proof of sufficiency}

The LU-invariance condition can be written as
\begin{equation}
F(\rho,\epsop)=0,
\end{equation}
where we define
\begin{equation}
F(\rho,\epsop)=f(e^{i\epsop}\rho e^{-i\epsop})-f(\rho)
\end{equation}
with $\epsop$ being a local hermitian operator. This condition has
to be satisfied for every $\rho$ and every $\epsop$. By expanding
up to first order in $\epsop$ we obtained condition
\eqref{mixed1}, which is equivalent to
\begin{equation}
\tr\left\{ \left.\frac{\partial F(\rho,\epsop)}{\partial\epsop}
  \right|_{\epsop=\hat{0}}\epsop\right\} = 0.
\end{equation}
This is a linear form of the components of $\epsop$ and the
requirement that it vanishes for every $\epsop$ implies that
\begin{equation}
\left.\frac{\partial F(\rho,\epsop)}{\partial\varepsilon_{ij}}
\right|_{\epsop=\hat{0}}=0.
\end{equation}
This has to be satisfied for every $\rho$. Consider the first
derivative of $F(\rho,\epsop)$ with respect to $\varepsilon_{ij}$,
taken at an arbitrary point $\epsop_0$. We have
\begin{equation}
\left.\frac{\partial F(\rho,\epsop)}{\partial\varepsilon_{ij}}
  \right|_{\epsop=\epsop_0}
= \left.\frac{\partial
F(\rho,\epsop_0+\epsop)}{\partial\varepsilon_{ij}}
  \right|_{\epsop=\hat{0}}.
\end{equation}
But from the form of $F(\rho,\epsop)$ one can see that
$F(\rho,\epsop_0+\epsop)=F(\rho',\epsop)$, where
$\rho'=e^{i\epsop_0}\rho e^{-i\epsop_0}$. Therefore

\begin{equation}
\left.\frac{\partial F(\rho,\epsop)}{\partial\varepsilon_{ij}}
  \right|_{\epsop=\epsop_0}
= \left.\frac{\partial F(\rho',\epsop)}{\partial\varepsilon_{ij}}
  \right|_{\epsop=\hat{0}} =0,
\end{equation}
i.e., the first derivatives of $F(\rho,\epsop)$ with respect to
the components of $\epsop$ vanish identically. This means that
$F(\rho,\epsop)=F(\rho,\hat{0})=0$ for every $\epsop$ and
condition \eqref{mixed1} is sufficient.

The condition for non-increase on average under local generalized
measurements \eqref{nonincrease4} can be written as
\begin{equation}
G(\rho,\epsop) \leq 0, \label{nonincrease5}
\end{equation}
where
\begin{equation}
G(\rho,\epsop) =p_1 f(\Mhat_1\rho \Mhat_1/p_1) + p_2 f(\Mhat_2\rho
\Mhat_2/p_2) - f(\rho).
\end{equation}
The operators $\Mhat_1$ and $\Mhat_2$ in terms of $\epsop$ are
given by \eqref{mm}, and the probabilities $p_1$ and $p_2$ are
defined as before. As we have argued in section III, it is
sufficient that this condition is satisfied for infinitesimal
$\epsop$. By expanding the condition up to second order in
$\epsop$ we obtained condition \eqref{mixed3}, which is equivalent
to
\begin{equation}
\tr\left\{
\left.\frac{\partial^2G(\rho,\epsop)}{\partial\epsop^{\otimes2}}
  \right|_{\epsop=\hat{0}}\epsop^{\otimes 2} \right\} \leq 0.
\end{equation}
Clearly, if this condition is satisfied by a strict inequality, it
is sufficient, since corrections of higher order in $\epsop$ can
be made arbitrarily smaller in magnitude by taking $\epsop$ small
enough. Concerns about the contribution of higher-order
corrections may arise only if the second-order correction to
$G(\rho, \epsop)$ vanishes in some open vicinity of $\rho$ and
some open vicinity of $\epsop$ (we have assumed that the function
$f(\rho)$ is continuous). But the second-order correction is a
real quadratic form of the components of $\epsop$ and it can
vanish in an open vicinity of $\epsop$, only if it vanishes for
every $\epsop$, i.e., if
\begin{equation}
\left.\frac{\partial^2G(\rho,\epsop)}{\partial\varepsilon_{ij}\partial\varepsilon_{kl}}
  \right|_{\epsop=\hat{0}}=0.
\label{zero}
\end{equation}
We will now show that if \eqref{zero} is satisfied in an open
vicinity of $\rho$, there exists an open vicinity of
$\epsop=\hat{0}$ in which all second derivatives of
$G(\rho,\epsop)$ with respect to $\epsop$ vanish identically. This
means that all higher-order corrections to $G(\rho,\epsop)$ vanish
in this vicinity and \eqref{nonincrease5} is satisfied with
equality.

Consider the two terms of $G(\rho,\epsop)$ that depend on
$\epsop$:
\begin{equation}
G_1(\rho,\epsop)=p_1 f(\Mhat_1\rho \Mhat_1/p_1),
\end{equation}
\begin{equation}
G_2(\rho,\epsop)=p_2 f(\Mhat_2\rho \Mhat_2/p_2).
\end{equation}
They differ only by the sign of $\epsop$, i.e. $G_1(\rho,\epsop)
=G_2(\rho,-\epsop)$, and therefore
\begin{equation}
\left.\frac{\partial^2
G_1(\rho,\epsop)}{\partial\varepsilon_{ij}\partial\varepsilon_{kl}}\right|_{\epsop=\hat{0}}=\left.\frac{\partial^2
G_2(\rho,\epsop)}{\partial\varepsilon_{ij}\partial\varepsilon_{kl}}\right|_{\epsop=\hat{0}}=\frac{1}{2}\left.\frac{\partial^2
G(\rho,\epsop)}{\partial\varepsilon_{ij}\partial\varepsilon_{kl}}\right|_{\epsop=\hat{0}}.
\end{equation}
If \eqref{zero} is satisfied in an open vicinity of $\rho$, we
have
\begin{equation}
\left.\frac{\partial^2
G_1(\rho,\epsop)}{\partial\varepsilon_{ij}\partial\varepsilon_{kl}}\right|_{\epsop=\hat{0}}=\left.\frac{\partial^2
G_2(\rho,\epsop)}{\partial\varepsilon_{ij}\partial\varepsilon_{kl}}\right|_{\epsop=\hat{0}}=0
\label{cond}
\end{equation}
in this vicinity. Consider the second derivatives of
$G(\rho,\epsop)$ with respect to the components of $\epsop$, taken
at a point $\epsop_0$:
\begin{equation}
\begin{split}
\left.\frac{\partial^2G(\rho,\epsop)}{\partial\varepsilon_{ij}\partial\varepsilon_{kl}}
  \right|_{\epsop=\epsop_0}
=
\left.\frac{\partial^2G_1(\rho,\epsop)}{\partial\varepsilon_{ij}\partial\varepsilon_{kl}}
  \right|_{\epsop=\epsop_0}
+
\left.\frac{\partial^2G_2(\rho,\epsop)}{\partial\varepsilon_{ij}\partial\varepsilon_{kl}}
  \right|_{\epsop=\epsop_0}  \\
=
\left.\frac{\partial^2G_1(\rho,\epsop_0+\epsop)}{\partial\varepsilon_{ij}\partial\varepsilon_{kl}}
  \right|_{\epsop=\hat{0}}
+
\left.\frac{\partial^2G_2(\rho,\epsop_0+\epsop)}{\partial\varepsilon_{ij}\partial\varepsilon_{kl}}
  \right|_{\epsop=\hat{0}}.
\end{split}
\end{equation}

From the expression for $G_1(\rho,\epsop)$ one can see that
$\epsop$ occurs in $G_1(\rho,\epsop)$ only in the combination
$\sqrt{\frac{\id-\epsop}{2}}\rho\sqrt{\frac{\id-\epsop}{2}}$. In
$G_1(\rho,\epsop_0+\epsop)$ it will appear only in
$\sqrt{\frac{\id-\epsop_0-\epsop}{2}}\rho\sqrt{\frac{\id-\epsop_0-\epsop}{2}}$.
But
\begin{equation}
\sqrt{\frac{\id-\epsop_0-\epsop}{2}}=\sqrt{\frac{\id-\epsop'}{2}}\sqrt{\id-\epsop_0},
\end{equation}
where
\begin{equation}
\epsop'=\epsop(\id-\epsop_0)^{-1}.
\end{equation}
So we can write
\begin{equation}
\sqrt{\frac{\id-\epsop_0-\epsop}{2}}\rho\sqrt{\frac{\id-\epsop_0-\epsop}{2}}
= p'\sqrt{\frac{\id-\epsop'}{2}}\rho'\sqrt{\frac{\id-\epsop'}{2}},
\end{equation}
where
\begin{equation}
\rho' = \left( \sqrt{\id-\epsop_0}\rho\sqrt{\id-\epsop_0}
\right)/p' \label{rhoprime}
\end{equation}
and
\begin{equation}
p'=\tr\left\{\sqrt{\id-\epsop_0}\rho\sqrt{\id-\epsop_0}\right\}.
\end{equation}
Then one can verify that
\begin{equation}
G_1(\rho,\epsop_0+\epsop)=p'G_1(\rho',\epsop').
\end{equation}
Similarly
\begin{equation}
G_2(\rho,\epsop_0+\epsop)=p''G_2(\rho'',\epsop''),
\end{equation}
where
\begin{equation}
\epsop''=\epsop(\id+\epsop_0)^{-1},
\end{equation}
\begin{equation}
\rho'' = \left(\sqrt{\id+\epsop_0}\rho\sqrt{\id+\epsop_0}
\right)/p'', \label{rhodoubleprime}
\end{equation}
\begin{equation}
p''=\tr\left\{\sqrt{\id+\epsop_0}\rho\sqrt{\id+\epsop_0}\right\}.
\end{equation}
Note that $\partial\varepsilon'_{pq}/\partial\varepsilon_{ij}$ and
$\partial\varepsilon''_{pq}/\partial\varepsilon_{ij}$ have no
dependence on $\epsop$.  Nor do $p'$ and $p''$.  Therefore we
obtain
\begin{equation}
\begin{split}
\left.\frac{\partial^2
G(\rho,\epsop)}{\partial\varepsilon_{ij}\partial\varepsilon_{kl}}\right|_{\epsop=\epsop_0}
=p'\left.\frac{\partial^2
G_1(\rho',\epsop')}{\partial\varepsilon_{ij}\partial\varepsilon_{kl}}\right|_{\epsop=\hat{0}}+
p''\left.\frac{\partial^2
G_2(\rho'',\epsop'')}{\partial\varepsilon_{ij}\partial\varepsilon_{kl}}\right|_{\epsop=\hat{0}} \\
=
\sum_{p,q,r,s}\frac{\partial\varepsilon'_{pq}}{\partial\varepsilon_{ij}}
\frac{\partial\varepsilon'_{rs}}{\partial\varepsilon_{kl}}
p'\left.\frac{\partial^2
G_1(\rho',\epsop')}{\partial\varepsilon'_{pq}\partial\varepsilon'_{rs}}\right|_{\epsop'=\hat{0}}
 +\\
\sum_{p,q,r,s}\frac{\partial\varepsilon''_{pq}}{\partial\varepsilon_{ij}}
\frac{\partial\varepsilon''_{rs}}{\partial\varepsilon_{kl}}
p''\left.\frac{\partial^2
G_2(\rho'',\epsop'')}{\partial\varepsilon''_{pq}\partial\varepsilon''_{rs}}\right|_{\epsop''=\hat{0}}.
\end{split}
\end{equation}
We assumed that \eqref{cond} is satisfied in an open vicinity of
$\rho$. If $\rho'$ and $\rho''$ are within this vicinity, the
above expression will vanish. But from \eqref{rhoprime} and
\eqref{rhodoubleprime} we see that as $\|\epsop_0\|$ tends to
zero, the quantities $\|\rho'-\rho\|$ and $\|\rho''-\rho\|$ also
tend to zero. Therefore there exists an open vicinity of
$\epsop_0=\hat{0}$, such that for every $\epsop_0$ in this
vicinity, the corresponding $\rho'$ and $\rho''$ will be within
the vicinity of $\rho$ for which \eqref{cond} is satisfied and
\begin{equation}
\left.\frac{\partial^2
G(\rho,\epsop)}{\partial\varepsilon_{ij}\partial\varepsilon_{kl}}\right|_{\epsop=\epsop_0}=0.
\end{equation}
This means that higher derivatives of $G(\rho,\epsop)$ with
respect to the components of $\epsop$ taken at points in this
vicinity will vanish, in particular derivatives taken at
$\epsop=\hat{0}$. So higher order corrections in $\epsop$ to
$G(\rho,\epsop)$ will also vanish. Therefore $G(\rho,\epsop)=0$ in
the vicinity of $\rho$ for which we assumed that \eqref{mixed3} is
satisfied with equality, which implies that condition
\eqref{mixed3} is sufficient.


\begin{thebibliography}{QCAE}

\bibitem{Bennett96a}  C. H. Bennett, H. J. Bernstein, S. Popescu, and B. Schumacher, Phys.~Rev.~A {\bf 53}, 2046 (1996).

\bibitem{Bennett96b} C. H. Bennett, G. Brassard, S. Popescu, B. Schumacher, J. A. Smolin, and W. K. Wootters, Phys.~Rev.~Lett. {\bf 76}, 722 (1996).

\bibitem{Bennett96c} C. H. Bennett, D. P. DiVincenzo, J. A. Smolin, and W. K. Wootters, Phys.~Rev.~A {\bf 54}, 3824 (1996).

\bibitem{Nielsen99} M. A. Nielsen, Phys.~Rev.~Lett. {\bf 83}, 436--439 (1999).

\bibitem{Vidal99} G. Vidal, Phys.~Rev.~Lett. {\bf 83}, 1046--1049 (1999).

\bibitem{Jonathan99a} D. Jonathan, and M.B. Plenio, Phys.~Rev.~Lett. {\bf 83}, 1455 (1999).

\bibitem{Hardy99} L. Hardy, Phys.~Rev.~A {\bf 60}, 1912 (1999).

\bibitem{Jonathan99b} D. Jonathan,  and M.B. Plenio, Phys.~Rev.~Lett. {\bf 83}, 3566 (1999).

\bibitem{Vidal00a} G. Vidal, D. Jonathan, and M. A. Nielsen, Phys.~Rev.~A {\bf 62}, 012304 (2000).

\bibitem{Bennett99} C. H. Bennett, D. P. DiVincenzo, C. A. Fuchs, T. Mor, E. Rains, P. W. Shor, J. A. Smolin, and W. K. Wootters, Phys.~Rev.~A {\bf 59}, 1070--1091 (1999).

\bibitem{Vidal00b}  G. Vidal, J. Mod. Optics {\bf 47}, 255 (2000).

\bibitem{Oreshkov05} O. Oreshkov, and T. A. Brun, Phys.~Rev.~Lett. {\bf 95}, 110409 (2005).

\bibitem{Kempe99} J. Kempe, Phys.~Rev.~A {\bf 60}, 910 (1999)

\bibitem{Brun04} T. A. Brun, Quantum Information and Computation {\bf 4}, 401 (2004).

\bibitem{Gingrich02} R. M. Gingrich, Phys.~Rev.~A {\bf 65}, 052302
(2002).

\bibitem{Acin00} A. Acin, A. Andrianov, L. Costa, E. Jane, J. I.
Latorre, and R. Tarrach, Phys.~Rev.~Lett. {\bf 85}, 1560 (2000)

\bibitem{Carteret00} H. Carteret, A. Higuchi, and A. Sudbery, J. Math.
Phys. {\bf 41}, 7932 (2000)

\bibitem{Sudbery01} A. Sudbery, J. Phys. A {\bf 34}, 643 (2001)

\bibitem{Coffman00} V. Coffman, J. Kundu, and W. K. Wooters, Phys.~Rev.~A {\bf 61},
052306 (2000).

\bibitem{Plenio05} M. B. Plenio, Phys.~Rev.~Lett. {\bf 95}, 090503 (2005)


\end{thebibliography}
\end{document}